\documentclass[a4paper,11pt]{article}
\pdfoutput=1 
\usepackage{amssymb}
\usepackage{bbm}
\usepackage{siunitx}
\usepackage{amsmath}
\usepackage{braket}
\usepackage{mathtools}
\usepackage[usenames,dvipsnames,svgnames,table]{xcolor}
\usepackage [utf8] {inputenc}
\usepackage{color}
\usepackage{subcaption}
\usepackage{lmodern}
\usepackage{footnote}
\usepackage{slashed}
 \usepackage{mathrsfs}
 \usepackage[pdftex] {graphicx}
\usepackage{multirow}
\usepackage{jheppub} 
\usepackage{here}
\usepackage{hyperref}
\usepackage[justification=centering, singlelinecheck=false]{caption}
\usepackage[list=true, labelfont=bf, labelformat=brace, position=top]{subcaption}

\newcommand{\eq}{\begin{eqnarray}}

\newcommand{\en}{\end{eqnarray}}

\title{Spurious poles in a finite volume}

\author[1]{Jin-Yi Pang,}
\affiliation[1]{College of Science, University of Shanghai for Science and Technology, Shanghai 200093, China}
\emailAdd{jypang@usst.edu.cn}

\author[2]{Martin Ebert,}
\affiliation[2]{Technische Universit\"at Darmstadt, Department of Physics,
64289 Darmstadt, Germany}
\emailAdd{mebert@theorie.ikp.physik.tu-darmstadt.de}

\author[2,3]{Hans-Werner Hammer,}
\emailAdd{hans-werner.hammer@physik.tu-darmstadt.de}
\affiliation[3]{ExtreMe Matter Institute EMMI and Helmholtz Forschungsakademie
  Hessen f\"ur FAIR (HFHF),
GSI Helmholtzzentrum für Schwerionenforschung GmbH,
64291 Darmstadt, Germany}

\author[4]{Fabian M\"uller,}
\affiliation[4]{Helmholtz-Institut f\"ur Strahlen- und Kernphysik (Theorie) and Bethe Center for Theoretical Physics, Universit\"at Bonn, 53115 Bonn, Germany}
\emailAdd{f.mueller@hiskp.uni-bonn.de}

\author[4,5]{Akaki Rusetsky,}
\affiliation[5]{Tbilisi State  University,  0186 Tbilisi, Georgia}
\emailAdd{rusetsky@hiskp.uni-bonn.de}

\author[6]{and Jia-Jun Wu}
\affiliation[6]{School of Physical Sciences, University of Chinese Academy of Sciences, Beijing 100049, China}
\emailAdd{wujiajun@ucas.ac.cn}

\abstract{
  \begin{sloppypar}
    \noindent
    Using effective-range expansion for the two-body amplitudes
   may generate spurious sub-threshold poles
   outside of the convergence range of
   the expansion. In the infinite volume, the emergence of such
   poles leads to the inconsistencies in
   the three-body equations, e.g., to the breakdown of unitarity.
   We investigate the effect of the spurious poles on the three-body
   quantization condition in a finite volume and show that it leads to a peculiar dependence
   of the energy levels on the box size $L$. Furthermore, within a
   simple model, it is
   demonstrated that the procedure for the removal of these
   poles, which was recently proposed in Ref.~\cite{Ebert:2021epn}
   in the infinite volume, can be adapted to the finite-volume calculations.
   The structure of the exact energy levels is reproduced with an accuracy
   that systematically improves order by order in the EFT expansion.
  \end{sloppypar}
}

\allowdisplaybreaks
\begin{document}
\maketitle
\flushbottom

\section{Introduction}

Recent years have seen a rapid progress
in the study of the three-body problem in a finite volume that is caused
by a necessity of analyzing lattice data in the three-particle sector
\cite{Kreuzer:2008bi,Kreuzer:2009jp,Kreuzer:2010ti,Kreuzer:2012sr,Briceno:2012rv,Polejaeva:2012ut,Jansen:2015lha,Hansen:2014eka,Hansen:2015zta,Hansen:2015zga,Hansen:2016fzj,Guo:2016fgl,Sharpe:2017jej,Guo:2017crd,Guo:2017ism,Meng:2017jgx,Briceno:2017tce,Hammer:2017uqm,Hammer:2017kms,Mai:2017bge,Guo:2018ibd,Guo:2018xbv,Klos:2018sen,Briceno:2018mlh,Briceno:2018aml,Mai:2019fba,Guo:2019ogp,Guo:2020spn,Blanton:2019igq,Pang:2019dfe,Jackura:2019bmu,Briceno:2019muc,Romero-Lopez:2019qrt,Konig:2020lzo,Brett:2021wyd,Hansen:2020zhy,Blanton:2020gha,Blanton:2020jnm,Pang:2020pkl,Hansen:2020otl,Romero-Lopez:2020rdq,Blanton:2020gmf,Muller:2020vtt,Blanton:2021mih,Muller:2021uur,Beane:2007es,Detmold:2008fn,Detmold:2008yn,Blanton:2019vdk,Horz:2019rrn,Culver:2019vvu,Fischer:2020jzp,Alexandru:2020xqf,Romero-Lopez:2018rcb,Blanton:2021llb,Mai:2021nul,Mai:2018djl}.
In particular, three conceptually equivalent formulations of the
three-body quantization condition (an equation that connects the
finite-volume energy spectrum with the infinite-volume observables in
the three-particle system) have been proposed -- the so-called
RFT~\cite{Hansen:2014eka, Hansen:2015zga},
NREFT~\cite{Hammer:2017uqm, Hammer:2017kms} and
FVU~\cite{Mai:2017bge,Mai:2018djl} approaches. A Lorentz-invariant
formulation of the NREFT approach was suggested
recently~\cite{Muller:2021uur}, and a three-body analog of the
Lellouch-L\"uscher formula, which relates the three-body decay amplitudes,
measured in a finite and in the infinite volume, has been
derived~\cite{Muller:2020wjo,Hansen:2021ofl}. For a more detailed overview,
we refer the reader to the two recent reviews on the
subject~\cite{Hansen:2019nir,Mai:2021lwb}.

All these approaches have in common that they connect the three-body
amplitudes in the infinite volume with the amplitudes in a finite
volume.  
The Faddeev equations for the three-body amplitude, and the par\-tic\-le-dimer
amplitudes obtained in the framework of effective field
theories with short-range interactions, contain the two-body
scattering amplitudes as an input. Furthermore, the integration over the
spectator momenta in the Faddeev equations extends to infinity.
Consequently, the energy variable in the two-body amplitude varies from
minus infinity to some threshold value determined by the total energy of
the three-particle system. In other words, in order to solve the equations,
the knowledge of the subthreshold two-body amplitude for arbitrarily
large negative energies, which is only weakly constrained
by the available two-body data in the threshold region, is required.
Based on the
decoupling of low- and high-energy degrees of freedom in a
field theory \cite{Appelquist:1974tg}, the low-energy three-body
observables should not
depend on the details of the two-body interaction
in the far away subthreshold region. Namely,
it should be possible to compensate any change of input from this region
by re-adjusting the local three-body coupling constants (often referred as
to the 'three-body force' in the context of the potential scattering theory),
leading to a consistent description of the
three-body problem in terms of only low-energy observables.

In practice, however, the situation is a bit more subtle. In the calculations,
one
has to use some
parameterization of the two-body amplitude, matched to the
low-energy two-body observables. The most prominent example of such
a parameterization is given by the effective-range expansion. Note
that this parameterization may become inconsistent for large negative
energies (this region,
 by definition, lies outside the range of applicability
of the effective-range expansion). Namely, if the effective-range
$r>0$, the two-body amplitude develops a pole at the distance
$k\sim 1/r$ below threshold (here, $k$ stands for the bound-state
momentum). Moreover, the residue of this pole is negative, which affects
the three-body unitarity {\em even in the low-energy region.}
This fact renders the application of the decoupling strategy
in the context of the present problem very intransparent. 
At this point, a conceptual difference to the two-body case is observed:
whereas in the two-body case the total center-of-mass (CM) energy is
a fixed variable and the low- and high-momentum regimes can be
separated without much ado, the three-body amplitude obeys an integral
equation, in which an integration is carried out over the CM energy
of the two-body system. Hence, the use of an
inconsistent parameterization for large momenta has an effect on the
low-momentum region of the three-body sector as well.

In the literature, one finds different prescriptions for `repairing' an
unphysical behavior of the two-body amplitude in the sub-threshold
region. The most straightforward way to do this consists in imposing an
upper cutoff on the spectator momentum in order to ensure that the
spurious singularities lie outside the integration range. This solution,
however, comes with a grain of salt, since the cutoff in this approach
cannot be made arbitrarily large and, moreover, the upper bound
on the cutoff depends on the order of the effective theory one is working.
All this complicates the use of the cutoff-dependence of the observables
for the error estimates in the effective-range expansion. Alternatively,
in Ref.~\cite{Bedaque:2002yg}, it was proposed to re-expand the full
two-body amplitude in the kernel of the integral equation, containing
the full dependence on the the scattering length $a$ and the
effective range $r$, in a power series in $r$.
This leads to a modification
of the two-body amplitude which can be made arbitrarily small within
the range of convergence of the effective-range expansion and,
at the same time, to a disappearance of the spurious subthreshold pole.
When expanded to linear order in $r$, all linear contributions in $r$
are included. However, some (small) higher-order contributions are included
as well, since the two-body amplitude appears in the kernel of the
integral equation.
This is avoided in a strictly perturbative approach where only
linear contributions in $r$ are kept at next-to-leading
order~\cite{Bedaque:1998km,Hammer:2001gh,Ji:2011qg}. This approach has been
extended to next-to-next-to leading order and works very well
phenomenologically~\cite{Ji:2012nj,Vanasse:2013sda}.
In Refs.~\cite{Platter:2006ev,Ryberg:2017tpv}, a slightly `less invasive'
strategy is adopted, namely, only the contribution from the spurious pole
is expanded, whereas the contribution from the physical (dimer) pole
is kept intact. A similar strategy is followed in the recent
work~\cite{Ebert:2021epn}.
A subtle difference between the two latter approaches consists in the fact
that the expansion in Ref.~\cite{Ebert:2021epn} is carried out in the two-body
energy variable, whereas in~\cite{Platter:2006ev,Ryberg:2017tpv} the
amplitudes are expanded in terms of
the relative momentum. Note that, only in the former case,
the expansion leads to a low-energy polynomial that can be
compensated by adjusting the renormalization prescription
in the three-body couplings.

We expect the fully perturbative approach of~\cite{Ji:2012nj,Vanasse:2013sda}
to be problematic numerically in a finite volume. In a finite box of size $L$,
the propagator of an S-wave  dimer is~\cite{Hammer:2017uqm,Hammer:2017kms}:
\eq\label{eq:tauL}
\tau_L({\bf k};{k^*}^2)=\frac{1}{k^*\cot\delta(k^*)+S({\bf k},{k^*}^2)}\, ,
\en
where ${\bf k},k^*$ denote the total three-momentum of the dimer and
the magnitude of the relative momentum of the two particles constituting
the dimer in their center-of-mass frame. Furthermore, $\delta(k^*)$ denotes
the pertinent phase shift and $S({\bf k},{k^*}^2)$ is given by the
sum\footnote{Note that this sum diverges and has to be properly regularized, e.g., by using dimensional regularization. The details can be found in Refs.~\cite{Hammer:2017uqm,Hammer:2017kms}.}
\eq\label{eq:S}
S({\bf k};{k^*}^2)=-\frac{4\pi}{L^3}\,\sum_{\bf p}\frac{1}{{\bf p}^2+{\bf p}{\bf k}+{\bf k}^2/4+{k^*}^2}\, ,\quad\quad {\bf p}=\frac{2\pi}{L}\,{\bf n}\, ,\quad
{\bf n}\in \mathbb{Z}^3\,.
\en
In the infinite volume, this sum turns into an integral, leading to a
well-known result.

The problem with expanding the finite-volume dimer propagator in a manner
proposed in
Refs.~\cite{Bedaque:1998km,Hammer:2001gh,Ji:2011qg,Ji:2012nj,Vanasse:2013sda}
is related to the singularities of the denominator. From
Eqs.~(\ref{eq:tauL}) and (\ref{eq:S})
it can be immediately seen that, in a finite volume,
the propagator has an infinite tower of poles above the elastic threshold,
corresponding to the finite-volume energy spectrum in the two-particle
subsystems. In the infinite volume, these poles condense and form an elastic
cut. In a finite volume,
the perturbative expansion in $r$ will not work in the vicinity of these
poles, producing denominators that become more and more singular.

The alternative scheme of Ref.~\cite{Ebert:2021epn} that
expands only the spurious pole was developed with this problem in mind.
The aim of the present paper is to apply this scheme and study the
consequences of the spurious pole in a finite volume. 
For example, it is known
that, in the infinite volume, spurious poles create problems with the
three-body unitarity. On the other hand, the finite-volume energy
spectrum is always real. How does the physical problem
manifest itself in a finite volume?
This and other questions will be answered below. In order to illustrate
the theoretical constructions with the help of numerical calculations, we
shall use a simple model with Yamaguchi potential as a reference theory (the
same as used in Ref.~\cite{Ebert:2021epn} in the infinite volume).

The layout of the present paper is the follows. In Sect.~\ref{sec:theory}
we give answer to the question, how the presence of a spurious pole affects the calculated finite-volume energy spectrum of the three-body system.
Here, we also formulate a systematic procedure for removing spurious poles that is an adaptation of the method of Ref.~\cite{Ebert:2021epn}
to the finite-volume calculations. In Sect.~\ref{sec:numerics}, it is shown that the method allows to reproduce the exact finite-volume spectrum in the model considered, while the precision improves order by order in the EFT expansion. Sect.~\ref{sec:concl} contains our conclusions.

\label{sec:intro}

\section{Removing spurious poles in the finite-volume calculations}
\label{sec:theory}

\subsection{The model}

As in Ref.~\cite{Ebert:2021epn}, we test our approach to remove the
spurious poles for a simple model,
whose energy levels  in a finite volume can be calculated numerically. 
We give a brief summary of the model but do not repeat all
formulae from~\cite{Ebert:2021epn}. The model features
identical spinless bosons with a mass $m=939.565~\mbox{MeV}$,
interacting pairwise through a
rank-one separable Yamaguchi potential in the S-wave:
\eq
V_Y(p,q)=\lambda\chi(p)\chi(q)\, ,\quad\quad
\chi(q)=\frac{\beta^2}{\beta^2+q^2}\, .
\en
The parameters of the model
$\lambda = -0.00013~\mbox{MeV}^{-2}$ and
$\beta = 278.8~\mbox{MeV}$
are chosen such that the two-body scattering
length $a=5.4194~\mbox{fm}$ and the effective range
$r=1.7536~\mbox{fm}$ are equal to those in the triplet
$np$ scattering channel.\footnote{Even if our model with bosons cannot
be put into one-to-one correspondence with the study of the real interactions
between the nucleons, analogies with the latter case might be still instructive.}
With this choice, the two-body amplitude has a
bound state pole on the real axis below threshold
at $E_2=-2.209~\mbox{MeV}$,
which is very close to the energy of the deuteron.
We shall refer to this bound state as the \textit{dimer}.

In the three-particle sector, we set the three-body force identically
to zero and choose an ultraviolet cutoff $\Lambda=1.5~\mbox{GeV}$ on the
spectator momentum. Strictly speaking, such a cutoff is not required
for the Yamaguchi model, and the cutoff $\Lambda$ can be taken to infinity.
In the field-theoretical language, this corresponds
to a particular choice of the renormalization prescription in the
three-particle sector. With this choice, there exist two three-particle
bound states: the shallow and the deep ones with the binding energies
$E_s=-2.356~\mbox{MeV}$ and $E_d=-24.797~\mbox{MeV}$.
The shallow three-body bound state is very close to the particle-dimer
threshold at $E=E_2$.

\begin{figure}[t]
  \begin{center}
    \begin{minipage}{0.45\columnwidth}
    \includegraphics*[width=6.4cm]{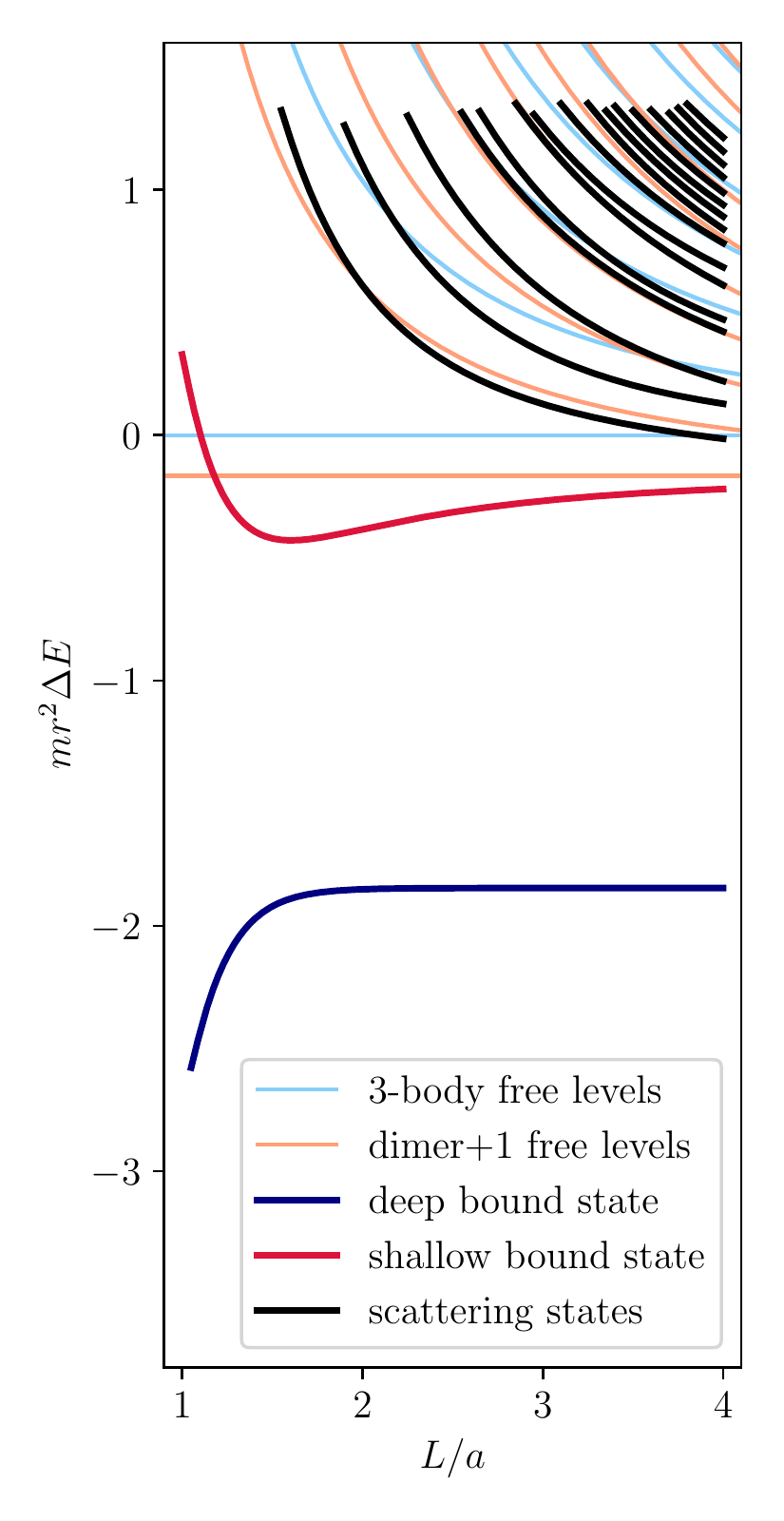}
\end{minipage}
    \begin{minipage}{0.45\columnwidth}
      \includegraphics*[width=6.4cm]{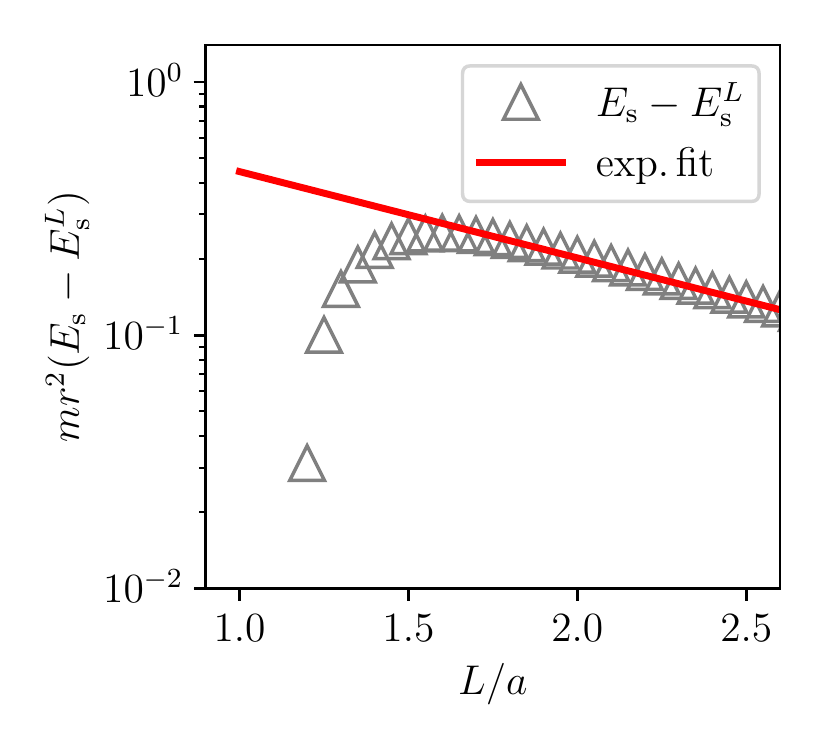}

 \vspace*{.8cm}     
      
       \includegraphics*[width=6.4cm]{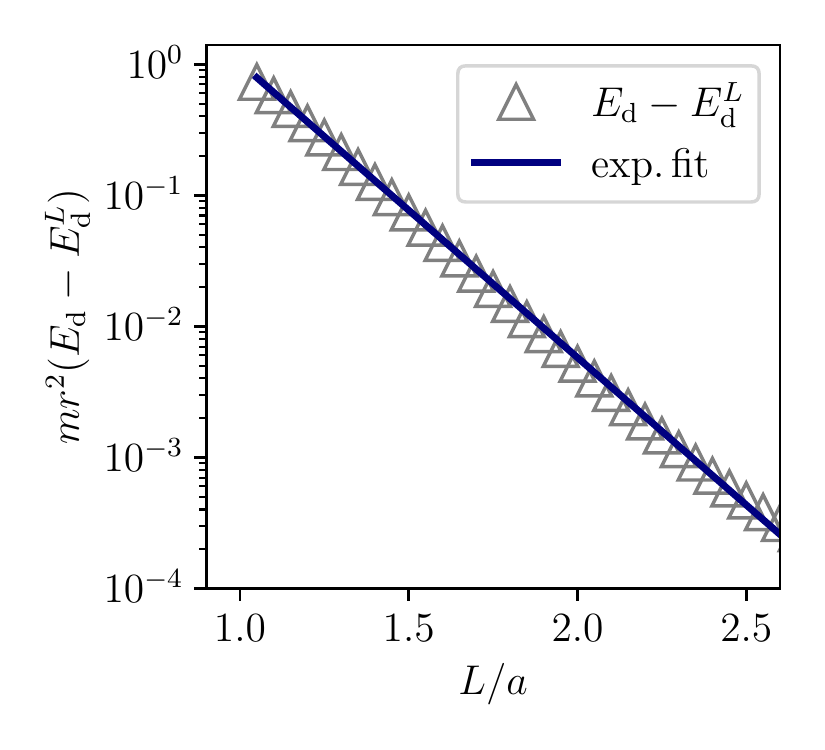}
 
\end{minipage}
 \end{center}
 \caption{Left panel: finite-volume three-particle
   spectrum of the Yamaguchi model for different values of $L$.
   Both scattering states and bound states are shown. For comparison,
   we also display the free three-particle and particle-dimer levels in
   a finite volume. Right panel: finite-volume correction to shallow and
   deep bound state energies. The results of the exponential fit are
   also shown. For explanation of symbols and lines see legend.
 }
\label{fig:yamaguchi}
\end{figure}

In order to obtain the spectrum in a finite box of size $L$,
in the scattering equations
one should replace the integrals over three-momenta by the sums over
discrete values ${\bf p}=2\pi{\bf n}/L$ and
${\bf n}\in \mathbb{Z}^3$. Then,
the poles of the Green functions in a finite volume determine the
discrete spectrum one is looking for (see Fig.~\ref{fig:yamaguchi}, left panel). Instead of the continuum, one gets now an
infinite
tower of discrete levels (scattering states)
that condense towards the lowest threshold,
as $L\to\infty$. Moreover, the lowest two levels converge to the deep
and the shallow bound state energies, respectively. As can seen from
Fig.~\ref{fig:yamaguchi}, for a large $L$, the finite-volume
correction to the spectrum decreases exponentially
$E_{s,d}-E_{s,d}^L\sim e^{-\kappa_{s,d} L}$, where $\kappa_s\simeq 26~\mbox{MeV}$ and
$\kappa_d\simeq 189~\mbox{MeV}$ from the fit.\footnote{It is well known
  that the finite-volume correction contains, in addition, a power-law
  dependence on $L$ in the prefactor~\cite{Meissner:2014dea,Hansen:2016ync,Konig:2017krd,Doring:2018xxx}. This effect, however, seems to be
  very suppressed in the data that correspond to a rather restricted
  interval in $L$.}

The results displayed in this section are standard and do not
deserve much attention. They are given here to provide the testing ground
for our method to remove spurious pole(s)~\cite{Ebert:2021epn}
within the non-relativistic effective field theory framework for the
three-body quantization condition in a finite
volume~\cite{Hammer:2017uqm, Hammer:2017kms}.
The discussion in this section sets stage for this test.

\subsection{Effective field theory}

The non-relativistic effective field theory that describes the bosonic Yamaguchi model in the two- and three-particle sectors, is defined by the following Lagrangian
\eq\label{eq:Lagrangian}
\mathscr{L}&=&\psi^\dagger\biggl(i\partial_0+\frac{\nabla^2}{2m}\biggr)\psi
-\frac{C_0}{2}\,(\psi^\dagger\psi)^2 
+\frac{C_2}{4}\,\biggl((\psi^\dagger{\stackrel{\leftrightarrow}{\nabla}}^2\psi^\dagger)\psi^2+\mbox{h.c.}\biggr)
\nonumber\\[2mm]
&-&\frac{D_0}{6}\,(\psi^\dagger\psi)^3
-\frac{D_2}{9}\,
\biggl((\psi^\dagger{\stackrel{\leftrightarrow}{\nabla}}^2\psi^\dagger)\psi^\dagger\psi^3+\mbox{h.c.}\biggr)+\cdots\, ,
\en
where $\psi$ is a non-relativistic field and
$\stackrel{\leftrightarrow}{\nabla}=\frac{1}{2}\,
(\stackrel{\rightarrow}{\nabla}-\stackrel{\leftarrow}{\nabla})$
denotes a Galilei-invariant derivative. The couplings $C_0,\,C_2$
can be related to the S-wave scattering length $a$ and the effective
range $r$, respectively (this relation takes a particularly simple form
in dimensional regularization). The S-wave two-body scattering
amplitude in the center-of-mass (CM) frame can be written as
\eq
\tau(k)=\frac{1}{k\cot\delta(k)-ik}\, ,
\en
where $k$ denotes the magnitude of the relative three-momentum
in the CM frame with energy $E=k^2/m$, and $\delta(k)$ is the scattering phase. The
effective-range expansion reads:
\eq
k\cot\delta(k)=-\frac{1}{a}+\frac{1}{2}\,rk^2+O(k^4)
=\frac{8\pi}{m}\,\frac{1}{-2C_0-2C_4k^2+O(k^4)}\, ,
\en
where dimensional regularization was used to regulate divergent loop
integrals.
The above equation defines the matching between the couplings $C_0,C_2$ and the effective-range expansion parameters.
The remaining couplings $D_0,D_2$ characterize the three-particle force
and will be determined from matching in the three-particle sector
(see below).

Working with the three-body systems, it is very convenient to use
the particle-dimer formalism~\cite{Kaplan:1996nv,Bedaque:1998kg,Bedaque:1998km,Bedaque:2002yg}. The dimers are auxiliary variables that are introduced
in the path integral. From this point of view, the formalism can be
used even if there exists no physical dimer. However, the formalism is particularly useful
if such a particle is present in the spectrum (as in the problem we are considering). In this case, the dominant feature
of the two-body amplitude is the formation of a close-by subthreshold pole.
Such a pole is described by a single diagram in the particle-dimer formalism and requires the resummation of an infinite tower of bubbles in the framework with no dimers present.

For the system we are investigating, the Lagrangian in the particle-dimer
formalism takes the form 
\eq\label{eq:Ldimer}
\mathscr{L}_d&=&\psi^\dagger\biggl(i\partial_0+\frac{\nabla^2}{2m}\biggr)\psi
+\sigma d^\dagger\biggl(i\partial_0+\frac{\nabla^2}{4m}+\Delta\biggr)d
\nonumber\\[2mm]
&+&\frac{f_0}{2}\,(d^\dagger\psi^2+\mbox{h.c.})+\cdots
+h_0d^\dagger d\psi^\dagger\psi +h_2d^\dagger d(\psi^\dagger\nabla^2\psi+(\nabla^2\psi^\dagger)\psi)+\cdots\, .
\en
Here, $d$ denotes the dimer field and ellipses stand either for the
higher-order terms in the derivative expansion, or for the contributions
from higher partial waves (here, only the contribution from the
S-wave is taken into account). Furthermore, the sign of $\sigma=\pm 1$ is
linked to the sign of the effective range -- in our case, $\sigma=-1$.
Finally, $\Delta,f_0,h_0,h_2,\ldots$ is just another set of parameters.
The requirement that both theories describe the same low-energy physics defines
the matching between these parameters. This matching has been
considered in the literature already many times
(see, e.g., Refs.~\cite{Bedaque:1999vb,Braaten:2004rn}), and will not
be discussed in detail here.

Furthermore, in the particle-dimer picture, the three-particle scattering
amplitude can be expressed algebraically via the (off-mass-shell)
particle-dimer scattering amplitude. This amplitude obeys the Faddeev
equation
\eq\label{eq:Faddeev}
M({\bf p},{\bf q};E)&=&Z({\bf p},{\bf q};E)+8\pi\int^\Lambda\frac{d^3{\bf k}}{(2\pi)^3}\,Z({\bf p},{\bf k};E)\tau(k^*)M({\bf k},{\bf q};E)\, ,
\en
where $E$ is the total energy of the particle-dimer system in the
CM frame, and $\tau(k^*)$ denotes the two-body amplitude. The energy
$E$ is assumed do have an infinitesimal positive imaginary part.
For our purposes, it will be convenient to choose the definition of the
relative momentum that is real below threshold. This can be achieved
by defining
\eq\label{eq:kstar}
k^*=\sqrt{\frac{3}{4}\,{\bf k}^2-mE}=ik\, ,
\en
and
\eq
\tau(k^*)=\biggl(k^*\cot\delta(k^*)+k^*\biggr)^{-1}\, .
\en
Finally, the driving term $Z$ is given by a sum of the one-particle exchange
term between the particle and the dimer, and a string of local couplings $H_0,H_2,\ldots$ which, {\em at the tree level,} can be related to the couplings
$h_0,h_2,\ldots$ from the particle-dimer Lagrangian:
\eq\label{eq:Z}
Z({\bf p},{\bf q};E)=\frac{1}{{\bf p}^2+{\bf q}^2+{\bf p}{\bf q}-mE}+\frac{H_0}{\Lambda^2}+\frac{3H_2}{8\Lambda^4}\,({\bf p}^2+{\bf q}^2)+\cdots\, .
\en
Note also that $H_0,H_2,\ldots$ should be dependent on the cutoff
$\Lambda$, in order to render the amplitude $M$ cutoff-independent
up to the higher-order terms. 

Furthermore, in Ref.~\cite{Bedaque:2002yg} is has been shown that,
introducing the trimer auxiliary field, it is possible to rewrite the Faddeev equation, replacing the momentum dependent term in $Z$, which is proportional to $H_2$, by a term that depends linearly on energy. The new driving term takes the form
\eq\label{eq:Zt}
\tilde Z({\bf p},{\bf q};E)=\frac{1}{{\bf p}^2+{\bf q}^2+{\bf p}{\bf q}-mE}+\frac{H_0}{\Lambda^2}
+\frac{\tilde H_2}{\Lambda^4}\,(mE+\gamma^2)+\cdots\, .
\en
The amplitude $M$ is the same in both cases, since the pertinent
second-order Lagrangians can be reduced to each other by the use
of the equations of motion and field redefinitions. Albeit
it remains to be seen,
whether such an equivalence can be extended to higher orders
as suggested in \cite{Griesshammer:2004pe}, this
issue is not of concern here because the discussion is restricted
to the second order. For this reason, we study {\em both} formulations,
referring to them as to the {\em $p$-scheme} and {\em $E$-scheme,} respectively.
We shall see that one indeed gets very similar numerical results, both
in the infinite as well as in a finite volume.

Next, we would like to address the matching of the couplings
$H_0$ and $H_2,\tilde H_2$. Most conveniently, this can be done by fixing
the dimer on mass shell and calculating the particle-dimer scattering
amplitude below the breakup threshold. One way would be to determine the two
couplings by matching the particle-dimer scattering length and the
effective range. For us it is more convenient
to match the particle-dimer scattering phase at two
values of the relative momentum in the CM system,
say, $p=10^{-3}~\mbox{MeV}$ and $p=10~\mbox{MeV}$.
Alternatively, instead of the matching point at
$p=10^{-3}~\mbox{MeV}$, one could use
the binding energy of the shallow
bound state. Both procedures yield practically identical results.

The quantization condition in a finite volume is obtained following
the standard path. The integral over the three-momentum ${\bf k}$
is replaced by a sum and, instead of $\tau(k^*)$, its finite-volume
counterpart appears, see Eqs.~(\ref{eq:tauL}) and (\ref{eq:S}).
The Faddeev equation becomes a system of linear equations that determines the finite-volume counterpart of the amplitude $M$ on the momentum grid. The zeros of the determinant of this equation define the position of the energy levels in a finite volume at a given value of the parameter $L$.

\subsection{Spurious poles}

It is well known that, if both the scattering length and the effective range are positive, the scattering amplitude $\tau(k^*)$ possesses a deep spurious pole on the real axis below threshold. This can be directly inferred from the explicit expression of the amplitude:
\eq
\tau(k^*)=\frac{1}{-1/a-r{k^*}^2/2+k^*}=\frac{-2/r}{(k^*-k_1)(k^*-k_2)}\, ,
\en
where
\eq
k_1=\frac{2/a}{1+\sqrt{1-2r/a}}\simeq \frac{1}{a}\, ,\quad\quad
k_2=\frac{1+\sqrt{1-2r/a}}{r}\simeq \frac{2}{r}\, .
\en
Close to the unitary limit $a\gg r$, the pole at $k_1$ corresponds to a shallow dimer state.
The second pole at $k_2$ is a spurious one. This can be verified immediately, since
the signs of the residua in two poles are opposite. Hence, the deep pole cannot correspond
to a physical state.

The existence of such an unphysical pole does not cause any problem, if one restricts
oneself exclusively to the two-body sector. In this sector, the total two-body energy
is an external variable and one can fix it above the unphysical pole. This is consistent,
because the pole at $k_2\sim r^{-1}$ emerges {\em outside the range of applicability of
  the effective-range expansion.} The situation is however totally different in the three-body
sector, where $\tau(k^*)$ appears in the integral equation (\ref{eq:Faddeev}), in which
the integral over the spectator momenta ${\bf k}$ is (formally) carried up to the infinity.
According to Eq.~(\ref{eq:kstar}), the variable $k^*$ then varies from $k^*=\sqrt{-mE-i\varepsilon}$ to $+\infty$ and thus hits the pole at $k^*=k_2$. This leads to problems
already in the infinite volume, related, for example, to the violation of unitarity
(caused by the wrong sign in the residue of the spurious pole) even at small three-body energies $E$. In other words, even
one is confident from the beginning that, owing to the decoupling theorem, the behavior at large spectator momenta should not alter the physical observables at low energies, the use of an inconsistent parameterization of the two-body amplitude in the high-energy region
may obscure this statement and render the use of the decoupling argument difficult.

In the context of the problem we are addressing in the present paper, it is interesting to find out, how the above-mentioned difficulty with the unitarity translates to a finite-volume setting. In a finite volume, the amplitudes are real, so the only place where this difficulty can manifest itself is the observation of a peculiar behavior of the the calculated energy levels.

What does such a peculiar behavior look like? In order to answer this question, consider first
a related phenomenon, namely the avoided level crossing. Assume that the
volume-dependent spectrum of a system is given by $E_1(L),E_2(L),\ldots$. In the vicinity
of an avoided level crossing, only two of them matter, and we neglect the rest. Assume
also that the interaction Lagrangian contains some parameter $g$, so that, if $g=0$, the states with the energies $E_1$ and $E_2$ do not interact with each other. An example is
provided by the decay of a kaon into two pions. If the weak decay constant tends to zero,
the one-kaon eigenstate of the total Hamiltonian decouples from
the two-pion scattering states and is almost volume-independent (up to exponential corrections).

Consider now a particular value of $L=\hat L$, for which $E_1(\hat L)=E_2(\hat L)$, if $g=0$. In the
vicinity of this $L$, the energy is determined from the secular equation:
\eq
&&\det\begin{pmatrix} (E-E_1) & gH_{12} \cr gH_{12} & (E-E_2) \end{pmatrix}=0\, ,
\nonumber\\[2mm]
\Longrightarrow\quad E&=&\frac{1}{2}\,(E_1+E_2)\pm\sqrt{\frac{1}{4}\,(E_1-E_2)^2+g^2H_{12}^2}\, .
\en
Here, $gH_{12}$ stands just for the matrix element of the Hamiltonian that describes
the transition between two states. Of course, if $g=0$, 
the two levels coincide at $L=\hat L$. It is now immediately seen that the levels {\em cannot} coincide, if $g\neq 0$, because the argument of the square root in the above equation never vanishes. If $g$ is small, they come very close to each other and then move apart again, as $L$ varies. This is how the avoided level crossing emerges.

In the problem we are studying, the spurious pole can be considered to correspond to some (fictitious) particle in the spectrum. An analog of the
free energy levels (at $g=0$) in the three-particle system is provided by non-interacting
three-particle levels, particle-dimer levels and the levels corresponding to a state with one particle and one fictitious particle. Furthermore, all
classes of levels have different $L$-dependence and condense towards pertinent thresholds, as $L\to\infty$. At some values of $L$, the free levels may cross. Consider, for example, the crossing of a level corresponding to one particle and one fictitious particle with any other level. The situation is similar as in the case of the avoided crossing, except that $g^2$ has now a different sign because of a different sign in the residue of the fictitious particle. Consequently, the levels may merge even for $g\neq 0$ -- the square root may vanish at some
discrete values of $L=L_c$. Moreover, the secular equation may have no real roots for some values of $L$ in the vicinity of $L=L_c$. This corresponds to the situation when two interacting levels merge and disappear.

\begin{figure}[t]
  \begin{center}
    \includegraphics*[width=9.cm]{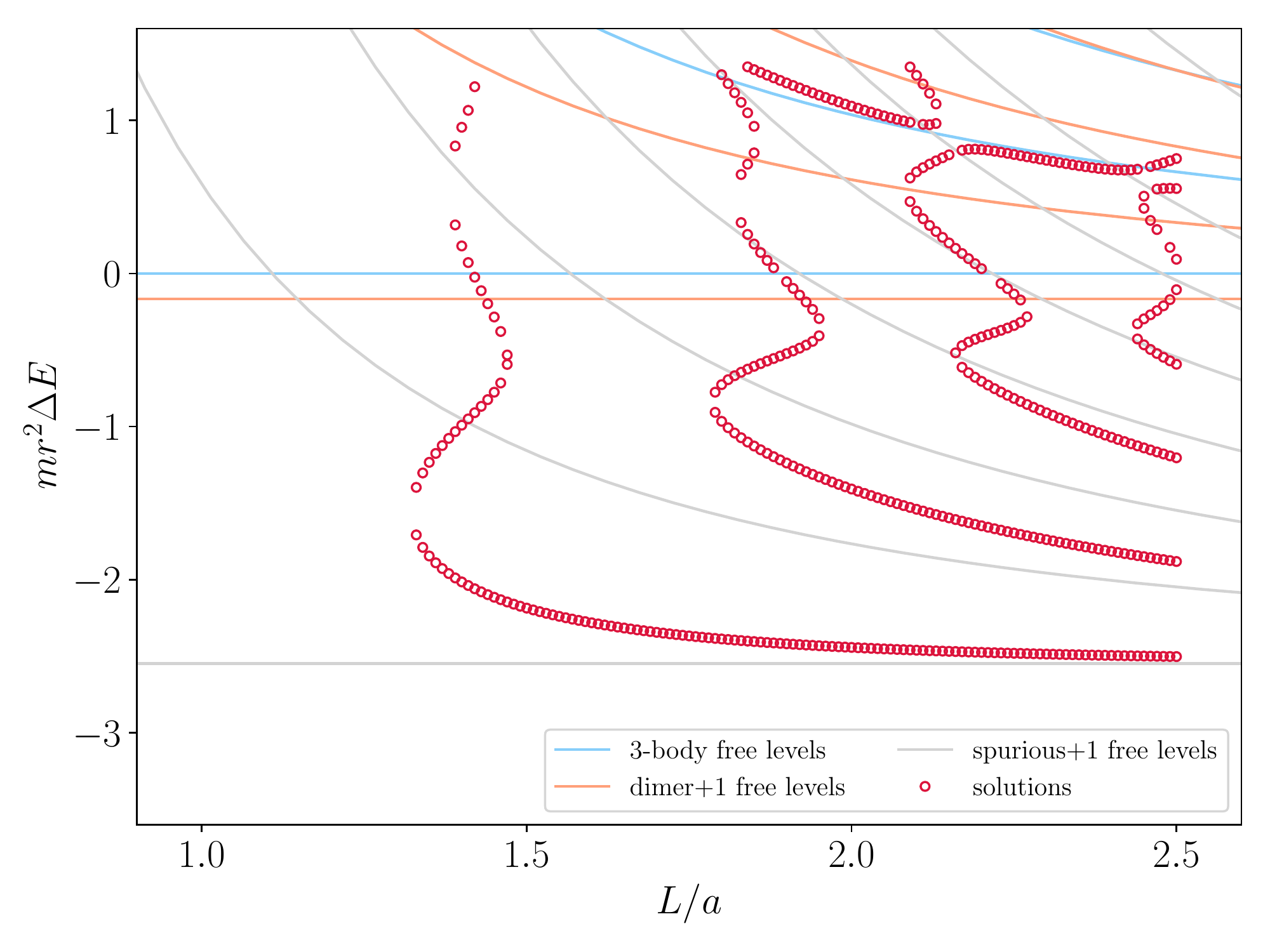}
    \caption{The solutions of the three-body quantization condition in the
      presence of  the spurious pole. One observes the unphysical merging and
      disappearance of energy levels
      that never occurs in a consistent theory.}
    \label{fig:spurious-levels}
  \end{center}
  \end{figure}

This behavior of the energy levels is verified trough the numerical solution
of the three-body quantization condition, (arbitrarily)
choosing $H_0=-0.414$ and $H_2=\tilde H_2=0$. The results
are shown in Fig.~\ref{fig:spurious-levels}. Note that such a behavior of the energy
levels cannot be observed in a consistent theory, where the spurious poles are absent.

\subsection{Removing the spurious pole}

In the literature, several approaches to the removal of the spurious pole are known. The most straightforward one consists in choosing a low enough cutoff $\Lambda$, so that the integration contour does not hit the singularity
at $k^*=k_2$. Obviously, the discussion of the cutoff-independence
within such an approach becomes extremely obscure. Alternatively,
since the problem with the spurious pole(s)
is absent at the leading order, one could refrain from summing
up the terms in the two-body amplitude that contain the effective range and higher-order parameters. This amounts up to an expansion
\eq
\tau(k^*)=\frac{1}{-1/a+k^*}+\frac{r{k^*}^2/2}{(-1/a+k^*)^2}+\cdots\, .
\en
Note that the different terms in this expansion count as $O(p^{-1})$, $O(p^0)$, $O(p)$ and so on (we have adopted the counting where the (unnaturally large) scattering length
$a=O(p^{-1})$, whereas $r$ and higher-order parameters count as $O(1)$).
Moreover, the spurious poles are gone, because the denominator $(-1/a+k^*)$ has
only one zero below threshold, corresponding to the shallow bound state.
The `perturbative' approach of
Refs.~\cite{Hammer:2001gh,Bedaque:1998km,Ji:2011qg,Ji:2012nj,Vanasse:2013sda} is
based exactly on this type of an expansion.

However, a direct generalization of this approach to the finite-volume calculations
could {\em potentially} encounter a problem. Namely, the quantity $(-1/a+k^*)$ has
no zeros
above threshold and the expansion is justified everywhere. On the contrary, the quantity
$-1/a+S({\bf k};{k^*}^2)$, which replaces it in a finite volume, has an infinite tower of roots
that correspond to the two-body energy levels. The perturbative expansion diverges
in the vicinity of these roots and thus cannot be applied everywhere above the two-body
threshold.
(see Eqs.~(\ref{eq:tauL}), (\ref{eq:S}) and the corresponding discussion
in Sect.~\ref{sec:intro}).

Thus, an alternative solution of the problem with the spurious poles is
desirable. Such an alternative was discussed in~\cite{Platter:2006ev,Ryberg:2017tpv} and, more recently, in~\cite{Ebert:2021epn}.
The latter prescription for the removal of the spurious pole has been shown
to amount to a renormalization of the local three-body
couplings~\cite{Ebert:2021epn}.\footnote{The following remark is in order. The three-body couplings, which are used for the renormalization, should be generally complex in order to account for the unitarity violation in the presence
of the spurious pole.} For this reason,
we shall further rely on the formulation given in Ref.~\cite{Ebert:2021epn} and adapt it to the finite-volume setting.

In brief, the method developed in~\cite{Ebert:2021epn} consists in the following.
First, in the infinite volume, a partial fraction decomposition is performed
\eq
\tau(k^*)=\frac{2(k_1+k_2)/r}{(k_2-k_1)(k^*+k_2)(k^*-k_1)}
-\frac{4k_2/r}{(k_2-k_1)({k^*}^2-k_2^2)}\, .
\en
Next, the quantity $\tau(k^*)$ is replaced by $\tau(k^*)-f(k^*)$, where
\eq\label{eq:subtractions}
f(k^*)=-\frac{4k_2/r}{(k_2-k_1)({k^*}^2-k_2^2)}
-\frac{4k_2/r}{(k_2-k_1)k_2^2}\biggl\{1+\frac{{k^*}^2}{k_2^2}+\frac{{k^*}^4}{k_2^4}+\cdots+O({k^*}^{(n+2)})\biggr\}\, .\quad\quad
\en
Note that $f(k^*)$ represents the contribution of the spurious pole minus the Taylor
expansion of the same quantity in powers of ${k^*}^2$. Furthermore, in
Ref.~\cite{Ebert:2021epn} it has been shown that the modification $\tau\to\tau-f$ amounts
to a change of the renormalization prescription.

Next, we discuss an analog of the above-described procedure in a finite volume. Using
a trial and error method, one could try to modify
the two-body amplitude (dimer propagator) in a finite volume as
\eq
\tau_L({\bf k};{k^*}^2)\to\tau_L({\bf k};{k^*}^2)-\bar f(k^*)\, ,
\en
where $\bar f(k^*)$, like $f(k^*)$, is a low-energy polynomial and independent of $L$
up to, possibly, exponential corrections which are not displayed explicitly. Furthermore,
beyond the two-particle threshold, the dimer propagator should coincide with its
infinite-volume counterpart. This guarantees, in particular, that
the spurious pole is gone in the modified propagator -- in the infinite, as well
as in a finite volume. Moreover, within the range of applicability of the effective theory,
where $k^*\ll k_2$, the initial and the modified propagators coincide up to
higher-order terms in the Taylor expansion.

The simplest choice is to assume that $\bar f(k^*)=f(k^*)$. Below, we shall demonstrate
that such a modification amounts to a change of the renormalization prescription
in a finite volume as well. In order to do this, let us recall that, in the infinite volume,
one has first rewritten the Faddeev equation as a system of two coupled equations
\eq
M({\bf p},{\bf q};E)&=&W({\bf p},{\bf q};E)
+8\pi\int^\Lambda\frac{d^3{\bf k}}{(2\pi)^3}\,W({\bf p},{\bf k};E)
(\tau(k^*)-\bar f(k^*))M({\bf k},{\bf q};E)\, ,
\nonumber\\[2mm]
W({\bf p},{\bf q};E)&=&Z({\bf p},{\bf q};E)
+8\pi\int^\Lambda\frac{d^3{\bf k}}{(2\pi)^3}\,Z({\bf p},{\bf k};E)\bar f(k^*)W({\bf k},{\bf q};E)
\nonumber\\[2mm]
&=&Z({\bf p},{\bf q};E)
+8\pi\int^\Lambda\frac{d^3{\bf k}}{(2\pi)^3}\,Z({\bf p},{\bf k};E)\bar f(k^*)Z({\bf k},{\bf q};E)+\cdots\, .
\en
In the infinite volume, the corrections in the effective potential $W$ that arise in the Born series represent low-energy polynomials (with complex coefficients, in general)~\cite{Ebert:2021epn}. Now, we wish to demonstrate
the same in a finite volume. For illustrative purpose, consider only the second iteration
and retain a single term that is proportional to $H_0^2$. This term has been discussed
in detail in Ref.~\cite{Ebert:2021epn}. A finite-volume counterpart of Eq.~(21) from that
paper is
\eq
I_{00}=\frac{1}{L^3}\sum_{\bf k}^\Lambda\biggl\{
\frac{1}{{k^*}^2-k_2^2}+\frac{1}{k_2^2}\,\biggl(1+\frac{{k^*}^2}{k_2^2}+\cdots\biggr)\biggr\}
=P+T\, .
\en
Here, $P$ and $T$ denote the pole term and the term coming from the Taylor-expanded part, respectively.
The second term is easy -- the sum over the low-energy polynomial is equal to the integral, up to the exponentially suppressed contributions. The pole term, however, is more subtle. Recalling the definition of ${k^*}^2$, it can be rewritten as
\eq
P=\frac{4}{3L^3}\,\sum_{\bf k}^\Lambda\frac{1}{{\bf k}^2-q_0^2}\, ,\quad\quad
q_0^2=\frac{4}{3}\,(k_2^2+mE)\, .
\en
It is immediately seen that, on the real energy axis, $P$ is {\em not} a low-energy polynomial -- indeed, it is proportional to the L\"uscher zeta-function which has an infinite tower
of singularities (note that the origin of the problem is the same as in the infinite volume -- namely, the singularity at ${\bf k}^2=q_0^2$ that gives rise
to the imaginary part in the infinite volume).
Hence, in order to expand this quantity, one has to move into the complex
energy- (complex $q_0^2$-) plane. Technically, it is most convenient to consider a complex
$q_0$-half-plane with $\mbox{Im}\,q_0>\varepsilon$. In the $q_0^2$-plane, this corresponds to a horizontally lying parabola
$\mbox{Im}\,(q_0^2)>2\varepsilon\sqrt{\mbox{Re}\,(q_0^2)+\varepsilon^2}$
which converges to the positive real axis in the limit $\varepsilon\to 0$. Using Poisson's
summation formula, we can transform the pole term into
\eq
P=P^\infty+\sum_{{\bf n}\neq 0}\frac{1}{3\pi nL}e^{inL\sqrt{q_0^2}}
=P^\infty+\sum_{{\bf n}\neq 0}\frac{1}{3\pi nL}e^{-nL\varepsilon}
\biggl\{1+\frac{iLn}{1!}\,(\sqrt{q_0^2}-i\varepsilon)+\cdots\biggr\}\, .\quad\quad
\en
Here, $P^\infty$ is the pole term in the infinite volume, which is
given in Ref.~\cite{Ebert:2021epn}, and the series are {\em convergent} for all
$\mbox{Im}\,q_0>\varepsilon$. Taking now into account that
\eq
q_0^2=\frac{2k_2}{3}\,\biggl\{1+\frac{mE}{2k_2^2}+\cdots\biggr\}\, ,
\en
one sees that the correction term is indeed a low-energy polynomial in the whole complex plane except a narrow strip along the positive real axis (the width of the strip is determined by the parameter $\varepsilon$). Moreover, the low-energy couplings are the same as in the infinite volume, up to exponential corrections that are proportional to
$e^{-\varepsilon L}$. Dropping these exponential terms, even in the limit $\varepsilon\to 0$,
is conceptually equivalent to dropping the imaginary part of the
effective potential in the infinite volume.\footnote{This is equivalent to dropping the low-energy polynomial and considering the limit $\varepsilon\to 0$ {\em afterwards}. } In other words, exactly
at this point the decoupling is imposed by hand, in order to repair the damage
that was caused by the use of the inconsistent parameterization containing
the spurious pole.

The general pattern is crystal clear from this simplest example. One can
always prove
that the corrections to the effective potential produce a low-energy polynomial in a complex plane everywhere except the vicinity of the real axis. Next, one discards exponential corrections and then takes the limit $\varepsilon\to 0$. As a result, one arrives at the prescription for removing the spurious pole(s), which was given above.

The above discussion, however, comes with a grain of salt, since it relies on perturbation theory. In the infinite volume, this does not create a problem. The situation is different in a finite volume, because the kernel $Z$ contains an exchange diagram between a particle
and the dimer. The latter is singular above the three-particle threshold, which implies that one should calculate $W$ from $Z$ non-perturbatively.
This may lead
to the distortion of the structure of the energy levels and thus
renders the replacement of
$W$ by $Z$ in a finite volume
questionable (In fact, we have explicitly checked that this
is exactly what happens: the singularity structure of $W$ and $Z$ is different).
In order to see, what goes wrong, note first that the quantity $\tau_L$
becomes zero exactly at those energies
where $Z$ is singular (i.e., at the free three-particle energies).
Consequently, the product $Z\tau_L$ in the
quantization condition is regular. On the contrary, $f$ is a continuous
function, so the product $Zf$ is singular at these energies. Now, writing down
explicitly the matrix equation that relates $W$ and $Z$ in a finite volume,
one can easily verify that the poles in $Z$, corresponding to the excited
levels, are split into several levels in $W$.  This splitting is very
small, as $f$ is small at small momenta. Hence, the singularity structure of
$W$ and $Z$ is indeed different and replacing $W$ by $Z$ in the quantization
condition cannot be justified.

This problem has a simple solution. Since the singularities emerge only above
the two-particle threshold, one can choose
\eq\label{eq:ffbar}
\bar f(k^*)=\left\{
  \begin{array}{l l}
    f(k^*)\, ,& {k^*}^2\geq 0\, ,\\[2mm]
    0\, ,& {k^*}^2<0\, .
  \end{array}
\right.
\en
In contrast to $f(k^*)$, this function is not a low-energy polynomial. However, it is
a low-energy polynomial up to the order one is working -- in fact, it vanishes up to this
order. This property suffices for our purposes, and using $\bar f(k^*)$ instead of $f(k^*)$ is fully
justified. On the other hand, $\bar f(k^*)$ vanishes in the area, where $Z$ becomes singular, and hence perturbation theory can be safely applied.
Obviously, this prescription is not unique and others can be designed.
But the one discussed here in the main text is perhaps the most
straightforward and transparent one. An example of an alternative scheme that works equally well
is discussed in Appendix~\ref{app:new-subtraction}.

We realize that part of the arguments given above are rather heuristic
and should be checked in practice.
In the next section, we shall explicitly verify
that this {\it ad hoc} prescription indeed enables us to relate the
finite-volume energy levels to coupling constants in the infinite
volume with a controlled accuracy. Thus it provides
an acceptable solution to the problem of spurious pole(s) in the context
of the finite-volume calculations.

\section{Numerical implementation and results}
\label{sec:numerics}

\subsection{Fixing of the effective couplings}

In order to carry out the calculation of the energy levels in a finite volume within the EFT,
matching of the effective three-body couplings is needed.
The calculations are done for leading order (LO), next-to-leading order (NLO) and next-to-next-to-leading order (N$^2$LO) in pionless EFT.
According to the standard power counting in the two- and three-body sectors, the following parameters appear at each order:
\begin{center}
\begin{tabular}{c|c|c}
  Order & 2-body parameters & 3-body parameters \\
  \hline
  LO & $a$ & $H_0$\\
NLO & $a,r$ & $H_0$\\
  N$^2$LO & $a,r$ & $H_0,H_2$\\
  \end{tabular}
\end{center}
This means, for example that, at LO, the scattering phase shift is given by $k\cot\delta(k)=-\dfrac{1}{a}$, and the kernel $Z$ contains only the non-derivative coupling proportional to $H_0$.

As mentioned earlier, the contact particle-dimer interaction can be written down in different forms. For instance, the derivative coupling $H_2$ may come together
with the three-momenta, see Eq.~(\ref{eq:Z}). Alternatively, the momentum dependence
can be traded in favor of the energy dependence, see Eq.~(\ref{eq:Zt}).
Hereafter, we shall refer to these alternatives as to the $p$-scheme and $E$-scheme,
respectively. The matching will be performed by setting the value of the particle-dimer
scattering phase shift at $p=10^{-3}~\mbox{MeV}$ and $p=50~\mbox{MeV}$ to the exact values, obtained in the Yamaguchi model.

\begin{figure}[t]
  \begin{center}
\includegraphics*[width=0.42\linewidth]{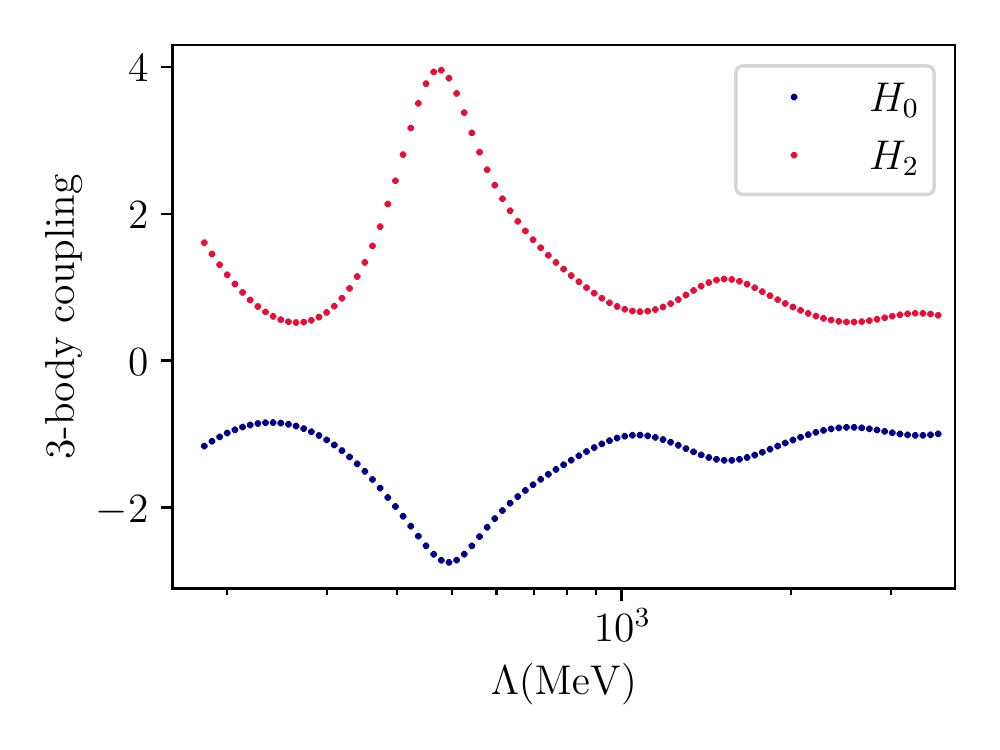}\quad
\includegraphics*[width=0.42\linewidth]{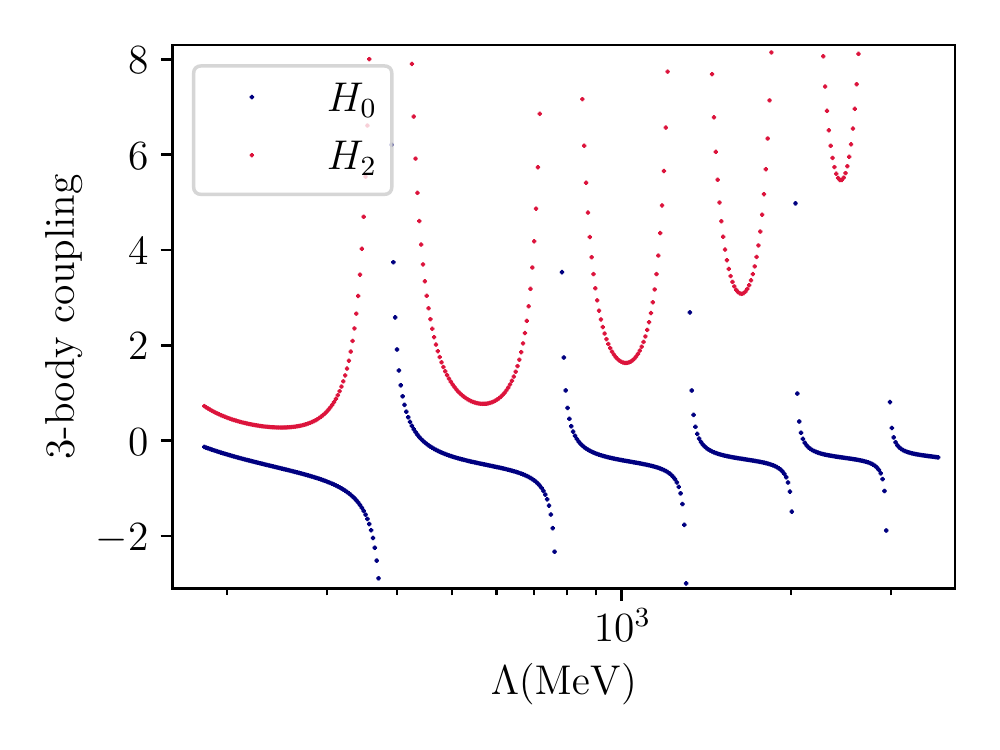}
\caption{The running three-body coupling constants $H_0$ and $H_2$
  as a function of the cutoff $\Lambda$: $p$-scheme (left panel) and $E$-scheme (right panel). The scheme dependence of the coupling constants is evident.}
\label{fig:H0-H2-running}
\end{center}
\end{figure}

Furthermore, the values of the couplings depend on the number of subtractions made, and
on the cutoff $\Lambda$. We shall perform two or three subtractions,
which corresponds to keeping two or three terms within the
curly brackets in Eq.~(\ref{eq:subtractions}), respectively.
In Fig.~\ref{fig:H0-H2-running}, we show this $\Lambda$-dependence for the $p$- and $E$-schemes in case of two subtractions.
As expected, this dependence is approximately log-periodic, both
for $H_0$ and for $H_2$.
Furthermore, in the $p$-scheme, in contrast to the $E$-scheme,
the dependence is smooth -- $H_0$ and $H_2$ never diverge.

The couplings, matched in this manner, can be used in the finite-volume
calculations.
Note that the choice of the ultraviolet cutoff $\Lambda$ is arbitrary.
All observables
are renormalization group invariant if the appropriate
running coupling constants are used in the calculation.
Thus different values of $\Lambda$ can be used for different values of
the box length $L$.
We shall use this freedom in the numerical implementation of the $p$-scheme and
will impose cutoff on the number of shells in the quantization condition,
rather than a fixed
cutoff in momentum space. Such a choice helps to eliminate small
numerical artifacts
in the energy levels, obtained through the solution of the quantization
condition.

\begin{table}[t]
  \begin{center}
    \begin{tabular}{|c|c|c|c|c|c|}
      \hline
 & 2-body parameters & subtr. & $\Lambda$ (GeV) & $H_0$ & $H_2$\tabularnewline
\hline 
\hline 
LO & $a$ & - & $2.2$ & $-2.006$ & -\tabularnewline
\hline 
NLO & \multirow{4}{*}{$a,r$} & \multirow{3}{*}{$2$} & $1.7$ & $-0.414$ & -\tabularnewline
\cline{1-1} \cline{4-6} 
N$^{2}$LO$_E$  &  & &$1.7$ & $-0.414$ & $3.401$ \tabularnewline
N$^{2}$LO$_p$ &  &  & $2.0\cdots 5.1$ & $-1.06\cdots -0.86$ & $0.47\cdots 0.71$\tabularnewline
\cline{1-1} \cline{3-6} 
N$^{2}$LO$_E$ &  & $3$ & $1.95$ & $-0.290$ & $3.948$\tabularnewline
\hline
\end{tabular}
\caption{The values of the parameters used in the finite-volume calculations.
 For the $p$-scheme the case of three subtractions was not considered.}
\label{tab:parameters}
\end{center}
\end{table}

In Table~\ref{tab:parameters}, we list the values of the parameters that were used in the finite-volume calculations. As mentioned above, in case of the
$p$-scheme we do not stick to a universal cutoff for all values of $L$.
In this scheme, the calculations were carried out for two subtractions only.

\subsection{Energy levels}

\begin{figure}[t]
  \begin{center}
\includegraphics*[width=10.cm]{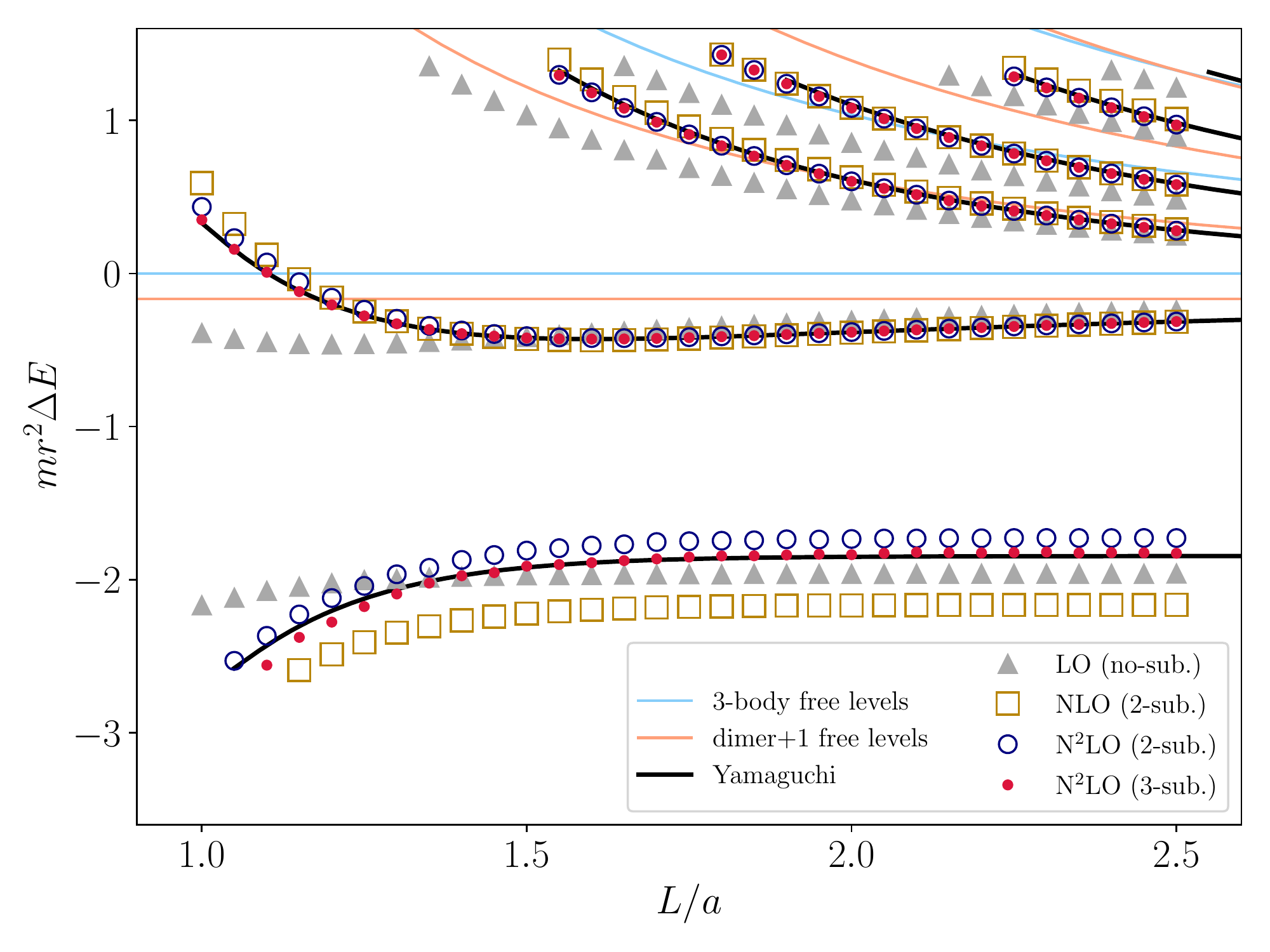}
\caption{Comparison of the EFT finite volume energy spectrum in the $E$-scheme
  at LO, NLO, and N$^2$LO with the Yamaguchi model. For convenience, the free
  3-particle and dimer-particle levels are also shown. For
  explanation of symbols and lines see legend.}
\label{fig:spectrum1}
\end{center}
\end{figure}

In Fig.~\ref{fig:spectrum1}, we display the calculated energy levels in
the $E$-scheme at different orders in the EFT.
The calculated levels agree very well with the result of the
Yamaguchi model and show
the expected convergence pattern from order to order.
The only exception is given by deep three-body state, which has a binding
energy of order $2/(mr^2)$ and thus is clearly outside the range of the EFT.
While the N$^2$LO result with three subtractions still appears to reproduce the
the energy from the Yamaguchi model rather well, the behavior of
different orders is not in agreement with the a priori expectation.
For  example, the NLO calculation is worse than the LO result.

Here, we refrain from a more detailed analysis of the convergence pattern using
Lepage plots, since such an analysis was already carried out in
Ref.~\cite{Ebert:2021epn}
and is performed in Appendix~\ref{app:new-subtraction} in the context of an
alternative subtraction scheme.
At N$^2$LO we show the EFT results for two and three subtractions in order to
highlight that performing more subtractions does
not necessarily lead to an increase of accuracy. An improvement is 
only obtained if the three-body coupling required to absorb the
resulting change in the renormalization prescription is present at the
given order~\cite{Ebert:2021epn}.

\begin{figure}[t]
  \begin{center}
\includegraphics*[width=10.cm]{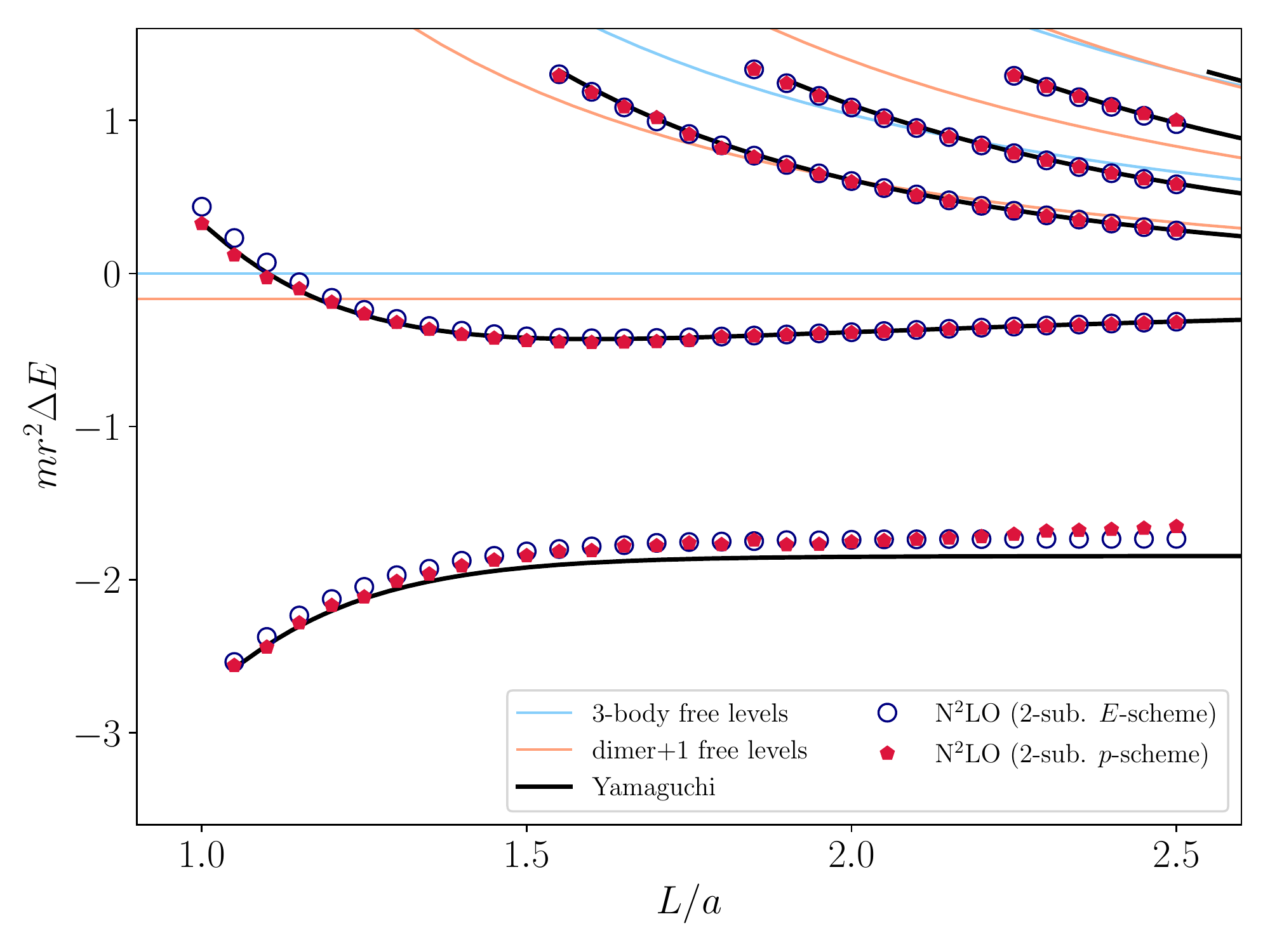}
\caption{Comparison of the EFT finite volume energy spectrum in the $p$- and $E$-schemes
  at N$^2$LO with 2 subtractions and the Yamaguchi results. For
  explanation of symbols and lines see legend}
\label{fig:spectrum2}
\end{center}
\end{figure}

In Fig.~\ref{fig:spectrum2}, a comparison of the EFT energy levels
in the $p$- and $E$-schemes
at N$^2$LO with two subtractions is given. The levels from both schemes
agree very well with each other and with the Yamaguchi result (except
for the deeply bound three-body state which is outside the convergence
range of the EFT). The two
schemes are equivalent in the infinite volume since
the pertinent Lagrangians are related through the use of the equations of
motion and field redefinitions~\cite{Bedaque:2002yg}.
Fig.~\ref{fig:spectrum2} demonstrates that the equivalence holds in a
finite volume as well. This is to be expected since the finite volume
only affects the infrared properties of the system.

To summarize, the generalization of the method to remove the spurious
pole from Ref.~\cite{Ebert:2021epn} to the finite-volume calculations works
well and thus can be used for the analysis of lattice data in the
three-particle sector.

\section{Conclusions}
\label{sec:concl}

In the three-body problem, the two-body scattering amplitude
enters at large negative energies up to the cutoff scale
of the the theory. Naively, one expects the low-energy three-body
physics to be insensitive to the behavior of the two-body amplitude
at large energies because of the decoupling of low- and high-energy
degrees of freedom in a field theory~\cite{Appelquist:1974tg}.
However, the two-body scattering amplitude
may develop unphysical singularities at energies that lie outside the
range of applicability of the given parameterization.
This happens for the effective range expansion starting
at linear order in the energy and, as a consequence, also
in short-range effective field theories that reproduce the effective range
expansion. In particular, the amplitude is known to develop a spurious
subthreshold pole at an energy of order $1/(mr^2)$ for $a,r>0$. This pole
is spurious since it has a residue with the  wrong sign and violates the
K\"all\'en-Lehmann spectral representation of the propagator.
When used as an input
in three-body calculations, the spurious poles lead to the breakdown of
unitarity {\em even at low energies,} thus jeopardizing the decoupling of
high- and low-momentum scales.

In the literature one finds different approaches which are designed to deal
with this problem, including partial resummations and perturbative expansions
in the range $r$, see Sect.~\ref{sec:intro}. In Ref.~\cite{Ebert:2021epn}, a new subtraction method
to remove the spurious poles was
proposed with finite volume applications in mind. 
It was further demonstrated that this prescription is equivalent
to a change of the renormalization prescription in the three-particle sector
(albeit the renormalized couplings turn out to be complex).

In the present paper, we studied the problem of spurious poles in a
finite volume and implemented the subtraction method from
Ref.~\cite{Ebert:2021epn}.
It was shown that the presence of such poles affects the
finite-volume spectrum in the three-particle system in a remarkable fashion.
Namely, the energy levels merge and disappear/appear at certain fixed values
of the box size $L$. This pathological behavior, which could be never observed in
a consistent quantum-mechanical system with a Hermitean Hamiltonian,
represents the finite-volume counterpart of the breakdown of unitarity,
observed in the infinite volume.

The main conclusions of our paper are as follows:

\begin{itemize}
  
\item[i)]
  A rather straightforward generalization of the approach of Ref.~\cite{Ebert:2021epn}
  to a finite-volume case is possible. We have discussed this generalization in detail and showed that the modification of the two-body amplitude boils down
  to a change of the renormalization prescription in the three-body sector
  also in the finite volume.

\item[ii)]
  We have verified our theoretical framework via explicit numerical calculations of
  the three-body spectrum, using a toy theory where the two-body interactions are described by a separable Yamaguchi potential. Within its region of
  applicability, the EFT expansion for the
  energy levels converges systematically to the exact result. 
  
\end{itemize}  

Thus, we conclude that the proposed approach to remove spurious poles
works well in a finite volume as well and can be readily used to analyze
lattice data in the three-particle sector. Further natural 
steps are the application to
analyze actual and/or mock lattice data  and the extension of our
scheme beyond N$^2$LO. The latter case presents some additional
theoretical challenges
because there maybe additional spurious poles and the applicability
of the $E$-scheme in this case has not been proven.

\begin{sloppypar}
{\em Acknowledgments:} We thank Harald Grie{\ss}hammer for useful discussions.
M.E. and H.-W.H. were supported by Deutsche
For\-schungs\-ge\-mein\-schaft  (DFG, German Research Foundation) --
  Project ID 279384907 -- SFB 1245.
M.E. in addition was supported by the Helmholtz Forschungsakademie Hessen
  für FAIR (HFHF).
  The work of F.M. and A.R. was  funded in part by
the Deutsche Forschungsgemeinschaft
(DFG, German Research Foundation) – Project-ID 196253076 – TRR 110.
A.R., in addition, thanks Volkswagenstiftung 
(grant no. 93562) and the Chinese Academy of Sciences (CAS) President's
International Fellowship Initiative (PIFI) (grant no. 2021VMB0007) for the
partial financial support.
J.J. Wu is supported by the
National Key R\&D Program of China under Contract No. 2020YFA0406400, and by the Key Research Program of the Chinese Academy of Sciences, Grant NO. XDPB15, and by National Natural Science Foundation of China under Grant No. 12175239. 
J.-Y. Pang is supported by National Natural Science Foundation of China under Grant No. 1213000064. 
\end{sloppypar}

\appendix

\section{An alternative subtraction scheme}
\label{app:new-subtraction}

As already mentioned in the text, the proposed subtraction scheme is not unique. In this
appendix we describe an alternative scheme, which can be used in the infinite as well as
in a finite volume. Conceptually, the new scheme is very simple and boils down to a
modification of the quantity $k^*\cot\delta(k^*)$ in the two-body amplitude {\em outside
  the range of applicability of the EFT.} This modification looks as follows
\eq\label{eq:phase-modified}
k^{*}\cot\delta(k^{*}) \to\left[-\frac{1}{a}-\frac{1}{2}rk^{*2}\right]
\left\{ 1+y\left(\frac{k^{*2}-k_{1}^{2}}{k_{2}^{2}}\right)^{2}\left(\frac{k^{*2}}{k_{2}^{2}}\right)^{2n}\right\}^{-1}\, .
\en
Here $y,n$ are two dimensionless parameters ($n\geq 1$ is an integer).
It is clear that
the first two terms in the expansion of the above expression in powers of ${k^*}^2$
have the same coefficients as the original effective-range expansion. Hence, the 
modified phase shift, given in Eq.~(\ref{eq:phase-modified}) is a valid option
for the continuation into the region beyond the range of applicability of the EFT
(this choice is at least as valid as the original effective-range expansion itself).
Now, one can use the freedom in the choice of the parameters $y$ and $n$ in order
to ensure that the spurious pole disappears. Choosing, for example, $n=1$, it can be
easily shown that the values of $y$ between
$y_{\min}=0.0337$ and $y_{\max}=1.54\cdot 10^5$ obey this requirement (One should
also note that, by construction, the pole in the amplitude at $k^*=k_1$ always stays put
and comes with the correct residue). In the following, we shall always stick to the case
$n=1$ (this quantity is an analog of the number of subtractions in the scheme, considered
in the main text). Concerning the choice of $y$, the naturalness
implies that the quantity $(k_1/k_2)^4y$ should be of order one. This
condition is in particular obeyed, if one tries to match the value of $y$ to the exact shape
parameter, obtained in the Yamaguchi model. The latter value, which lies within
the allowed interval for $y$, does not however lead to a reasonable
finite-volume spectrum, which is rather sensitive to the choice of $y$.
By using the trial and error method, we have verified
that the values around $y_{\min}$ that could be considered as a ``least
invasive subtraction,'' yield the best results. In the following, we shall
follow this finding and use the values of $y$
close to the lower end of the allowed interval.

\begin{figure}[t]
\begin{centering}
\includegraphics[width=0.6\linewidth]{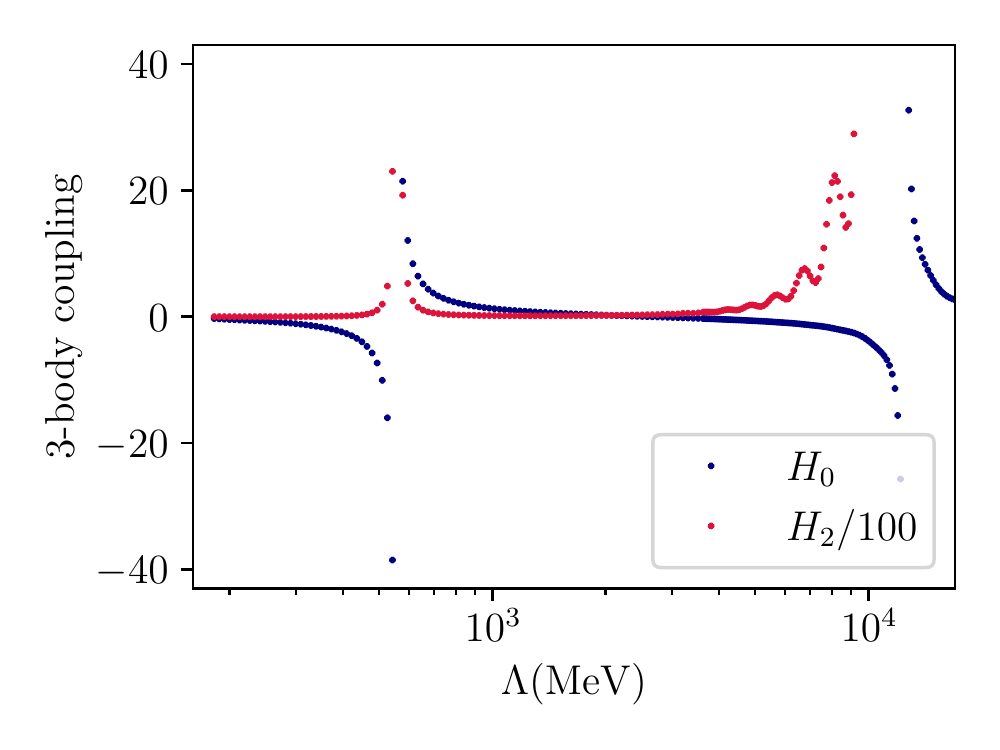}
\par\end{centering}
\caption{Running of the couplings $H_{0}$ and $H_{2}$ in the new scheme.}
\label{fig:sub2_running}
\end{figure}

At the next step, one has to perform the matching of the couplings $H_0,H_2$ for different
values of the cutoff $\Lambda$ (the $E$-scheme is exclusively
used below for the construction of the particle-dimer potential).
We again do this, matching the particle-dimer scattering
phase at two different values of the magnitude of the momentum:
at $p=10^{-3}~\mbox{MeV}$ and $p=50~\mbox{MeV}$. The matching at
$p=10~\mbox{MeV}$ has been also done and yields practically the same result.
The running of the couplings with $\Lambda$ is shown in Fig.~\ref{fig:sub2_running}.
It is seen that this running, as in other schemes, exhibits in general
a log-periodic behavior, albeit the detailed shape of the curves is of course different.
Note also that the model describes the second (deep) three-particle
bound state in the Yamaguchi model (which lies already
outside the region of the applicability
of the EFT) pretty well. For example, 
taking $y=0.05$ and choosing
$\Lambda$ as $1.5~\mbox{GeV}$, $1.7~\mbox{GeV}$ or $1.95~\mbox{GeV}$,
one gets $-24.2749~\mbox{MeV}$, $-24.5364~\mbox{MeV}$ and $-24.6055~\mbox{MeV}$ for the binding energy of the deep state, respectively (the exact result in the Yamaguchi
model is equal to $-24.7965~\mbox{MeV}$).

\begin{figure}[t]
  \begin{center}
\includegraphics*[width=0.45\linewidth]{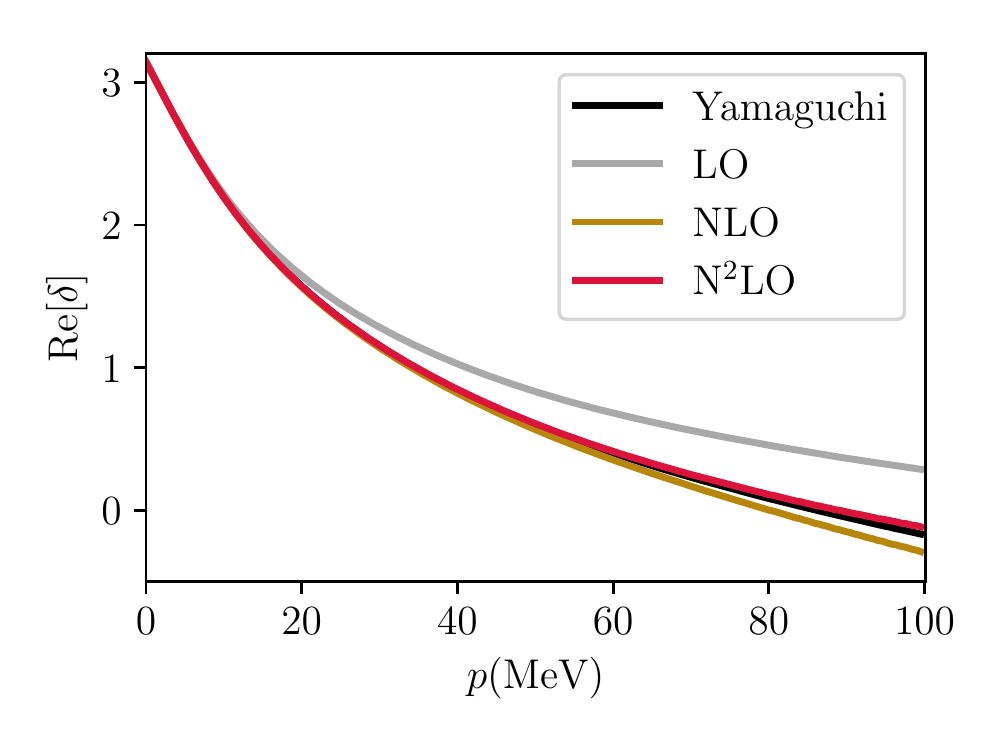}
\includegraphics*[width=0.45\linewidth]{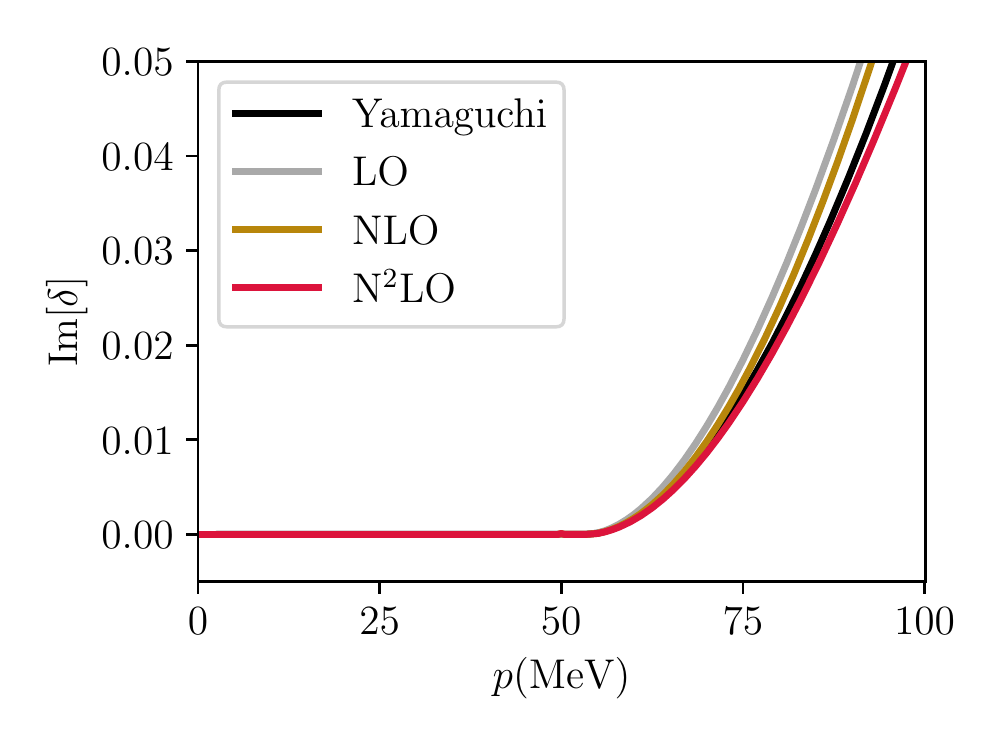}
\caption{The real and imaginary parts of the particle-dimer phase shift
at different orders of EFT. The values of parameters are $\Lambda=1.95\,\mbox{GeV}$ and $y=0.05$.
The number of subtractions (whenever needed) is chosen to be equal to one.}
\label{fig:Phaseshift_sub2}
\end{center}
\end{figure}

In Fig.~\ref{fig:Phaseshift_sub2}, we show the test of the alternative subtraction
method in the infinite volume. The value of the
cutoff $\Lambda=1.95\,\mbox{GeV}$ has been used in the calculations
(the same as later, in the calculations of the finite-volume spectrum), whereas
the value of $y$ was fixed at $y=0.05$.
The number of subtractions was taken equal to one. As seen, the convergence
of the EFT expansion is very good, and the result at N$^2$LO is almost in a
perfect agreement with the exact result, obtained in the Yamaguchi model.

A more detailed test of the convergence is provided by the so-called
Lepage plot and the consistency assessment.
In brief, in the Lepage plot, the EFT solution at different orders
is compared with the exact solution, whereas the consistency assessment implies
a comparison of the EFT solutions at two different values of $\Lambda$
(see Ref.~\cite{Ebert:2021epn} for more details).
Note also that the value of the cutoff chosen above is too high for carrying out
the consistency assessment, albeit performing calculations of observables
with any value
of cutoff is perfectly legitimate. For this reason, only to carry out
the check, we use lower values of the cutoff $\Lambda=0.25\,\mbox{GeV}$ and
$\Lambda=0.6\,\mbox{GeV}$. The parameter $y$ is again set to $y=0.05$.
The difference is fitted by a straight line
within the so-called opportunity window, and the slope gives the estimate
for the power of the terms that are neglected in the calculations.
As seen, both tests work very well, with the slope that consistently increases
roughly by one unit order by order.

\begin{figure}[t]
  \begin{center}
\includegraphics*[width=0.45\linewidth]{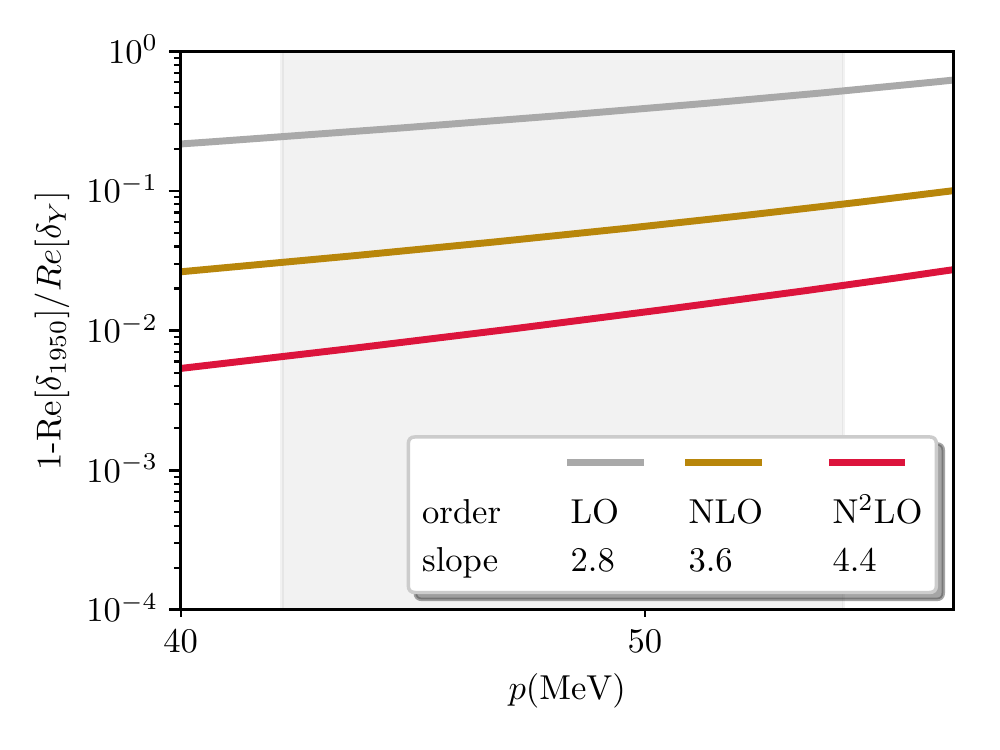}
\includegraphics*[width=0.45\linewidth]{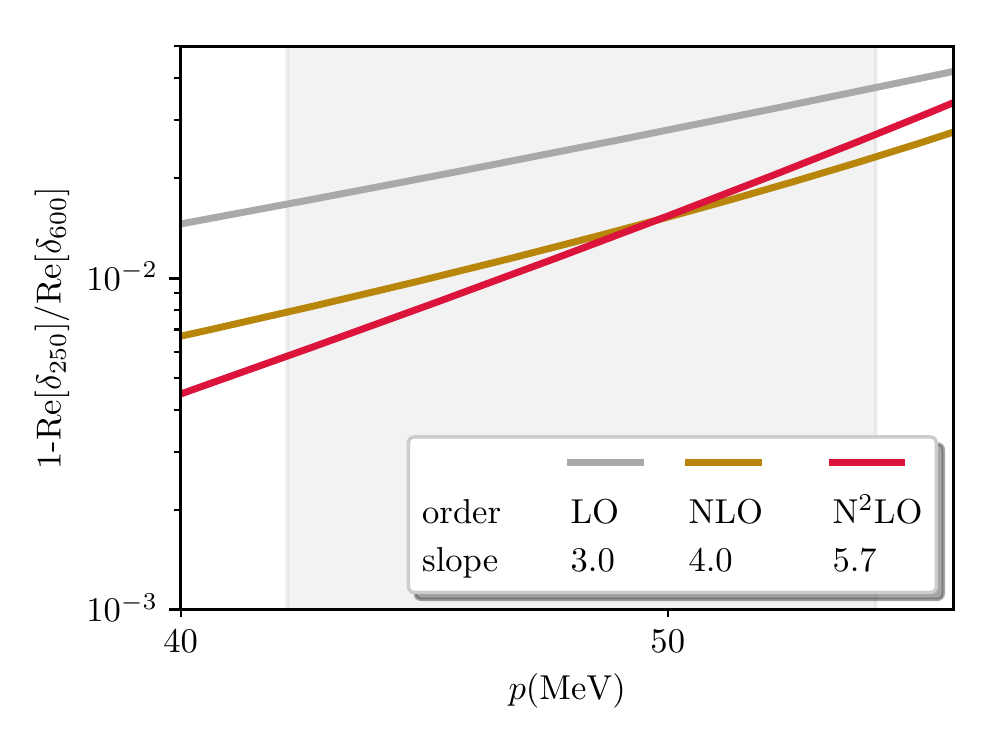}
\caption{The Lepage plot (left) and the consistency assessment (right).
The number of subtractions at NLO and N$^2$LO is equal to one. The
curves are fitted by straight lines within the opportunity window (shown
by the shaded area). The slopes are seen to increase order by order.
Note that the subscript of the phase shift indicates the value of the cutoff $\Lambda$,
used in the calculations. For example, $\delta_{250}$ denotes the phase shift
calculated using $\Lambda=250\,\mbox{MeV}$.}
\label{fig:Lepage_sub2}
\end{center}
\end{figure}

\begin{figure}[t]
\begin{centering}
\includegraphics[width=10.cm]{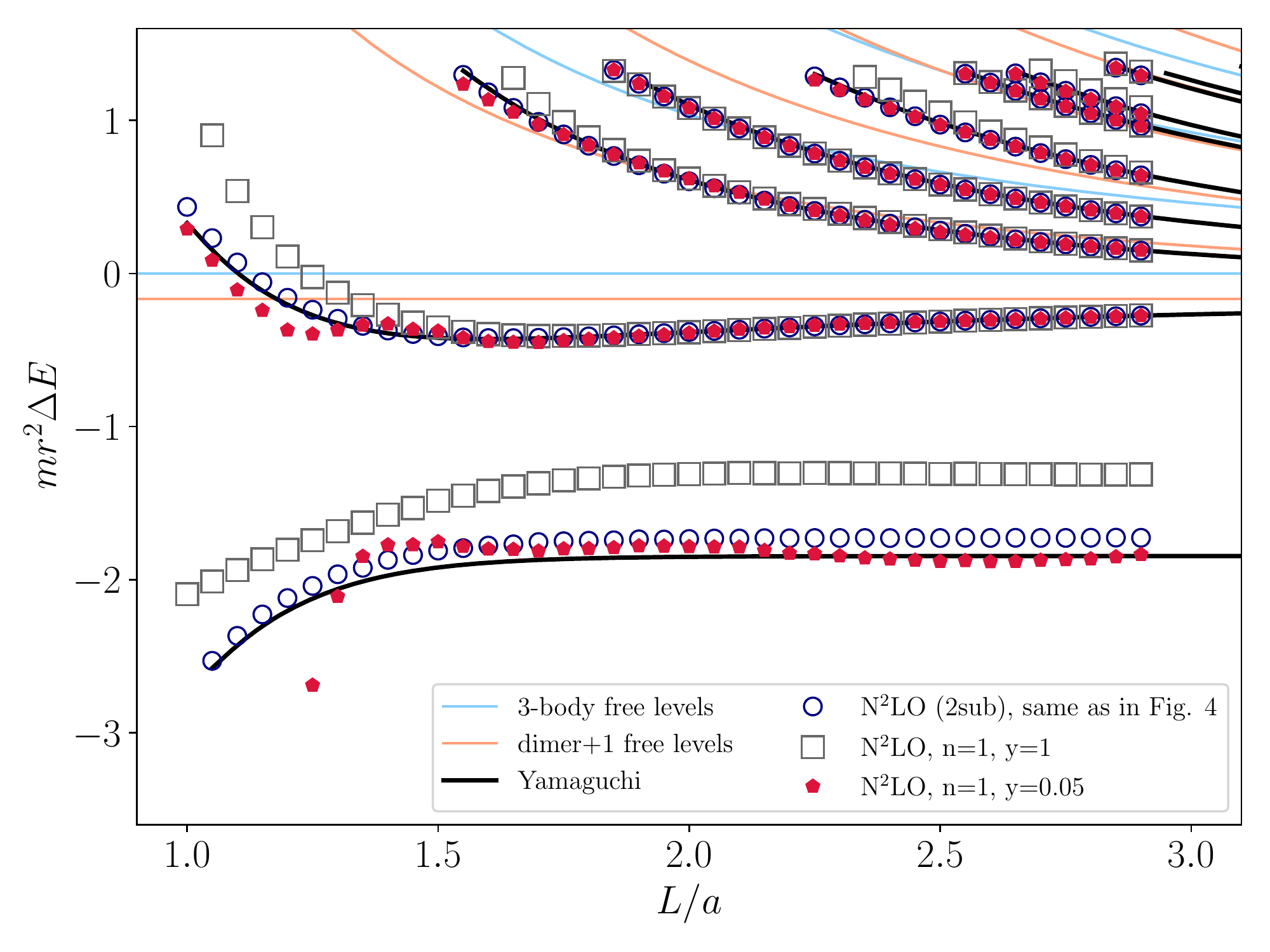}
\par\end{centering}
\caption{The finite volume spectrum, obtained with the use of the alternative subtraction scheme. For comparison, we show the result obtained in the scheme that is considered in the main text.}
\label{fig:sub2_spectrum}
\end{figure}

Finally, we present the calculation of the spectrum in a finite volume. The parameters
used in the calculations are given in Table~\ref{tab:pars}.
\begin{table}[H]
\begin{centering}
\begin{tabular}{c|c|c|c}
parameters & $\Lambda~\mbox{GeV}$ & $H_{0}$ & $H_{2}$\tabularnewline
\hline 
$n=1,~y=1$ & $1.95$ & $0.898$ & $-123$\tabularnewline
$n=1,~y=0.05$ &  & $0.273$ & $19.9$\tabularnewline
\end{tabular}
\par\end{centering}
\caption{The values of the parameters, used in the calculations of the finite-volume spectrum.}
\label{tab:pars}
\end{table}
The energy levels are shown in Fig.~\ref{fig:sub2_spectrum}. As seen, the quality of the
description of the levels in the vicinity of the threshold and above is comparable with
the case of the subtraction scheme, considered in the main text and is very good. The
description of the deep bound state is less impressive and strongly depends on the
chosen value for the parameter $y$. This however does not come at a big surprise, since
the deep bound state lies already outside the range of the applicability of the theory.

The advantage of the alternative subtraction scheme, considered in this appendix, lies in its
transparency. In addition, it is relatively easy to implement. However, a rather strong
dependence on the choice of the parameter $y$ (and, eventually, $n$), which was
observed above, could be considered as a moderate disadvantage. Namely, in the absence of the exact solution which one could compare with, a choice of the optimal value of $y$ may represent a problem. In addition, going to higher orders in EFT, $n=1$ is no more an option and
one has to choose a higher value for this parameter.

\bibliographystyle{unsrt}
\bibliography{references}

\end{document}